# Gravitational field around blackhole induces photonic spin-orbit interaction that twists light

Deng Pan, Hong-Xing Xu[‡]

*School of Physics and Technology, Wuhan University, Wuhan 430072, China*

*Corresponding authors. E-mail: ‡hxxu@whu.edu.cn*

The spin-orbit interaction (SOI) of light has been intensively studied in nanophotonics because it enables sensitive control of photons' spin degree of freedom and thereby the trajectory of the photons, which is useful for applications such as signal encoding and routing. A recent study [1] showed that the SOI of photons manifests in the presence of a gradient in the permittivity of the medium through which the photons propagate; this enhances the scattering of circularly polarized light and results in the photons propagating along twisted trajectories. Here we theoretically predict that, because of the equivalence between an inhomogeneous dielectric medium and a gravitational field demonstrated in transformation optics, a significant SOI is induced onto circularly polarized light passing by the gravitational lens of a black hole. This leads to: i) the photons to propagate along chiral trajectories if the size of the blackhole is smaller than the wavelength of the incident photons, ii) the resulting image of the gravitational lens to manifest an azimuthal rotation because of these chiral trajectories. The findings open for a way to probe for and discover subwavelength-size black holes using circularly polarized light.

# 1 Introduction

The spin-orbit interaction (SOI) is a fundamental phenomenon for the realization of various exotic spin dynamics and matters of nontrivial topological phases [2, 3]. The SOI also occurs for photons when circularly polarized light propagates in an inhomogeneous dielectric medium. The inhomogeneity of the medium forces photons to move along curved trajectories and couples their spin angular momentum (AM) with their orbital AM because of the conservation of total AM [4]. As a result of the SOI, circularly polarized photons acquire a transverse component of anomalous velocity, which shifts the trajectories of the light beam away from those given by ordinary geometric optics. The SOI-induced spin-dependent motion can be used to realize the splitting and routing of photons with different spins; recently, this has been intensively investigated in nanophotonics [5-13].

It has been shown that the SOI in light results in the presence of a gradient in permittivity of the medium [1]. Recent developments in transformation optics [14, 15] have also demonstrated the equivalence between an inhomogeneous dielectric medium and a gravitational field described in by the theory of general relativity. This equivalence has been utilized to mimic celestial phenomena such as black hole [16, 17] and gravitational lensing [18, 19] using photonic nanostructures. Conversely, it was also predicted that optical phenomena such as the SOI of light, which can be observed in the photonic nanostructures, can be realized in the presence of a gravitational field [20]. Motivated by these advancements we wondered how circularly polarized light would propagate through the gravitational lens of black holes that have a size similar or smaller to the wavelength of the incident light.

Here we show that the gravitational lens of a sub-wavelength black hole produces a significant SOI effect that leads to chiral trajectories of incident circularly polarized photons and a resulting rotation of the image of the black hole obtained by these photons. The findings predict SOI-induced signatures in circularly polarized light could be used to probe for and discover small blackholes.

## 2 Result

The motion of circularly polarized light in the gravitational field of a black hole is described by the equations [20]

$$\dot{\mathbf{p}} = -2GM \frac{\mathbf{r}}{cr^3} p \tag{1}$$

$$\dot{\mathbf{r}} = c \frac{\mathbf{p}}{p} F + \lambda \frac{2GM}{cr^3} \frac{\mathbf{r} \times \mathbf{p}}{p^2} \tag{2}$$

where $r$ and $p$ are the position and momentum vectors of the photons in the light beam, $\lambda = \pm \hbar$ is the helicity of the photons, and $F = V/W$ is derived from the Schwarzschild metric of the black hole $ds^2 = V^2 (ct)^2 - W^2 (d\mathbf{r} \cdot d\mathbf{r})$, with [21]

$$V = \left(1 - \frac{GM}{2c^2 r}\right)\left(1 + \frac{GM}{2c^2 r}\right)^{-1} \qquad W = \left(1 + \frac{GM}{2c^2 r}\right)^2. \tag{3}$$

Eq. (1) and the first term on the right of Eq. (2) show the ordinary gravitational lens effect, where the deviation of the momentum vectors is along the radial direction. More interestingly, for circularly polarized light, the second term of Eq. (2) introduces a component that is perpendicular to its momentum and the position vector. This anomalous velocity component depends on the helicity of the photon, which can lead to the separation of photons with opposite helicities. This phenomenon is identical to the scattering by a nanosphere [1], in which case the gradient of the refractive index induces the SOI during the scattering. Here the SOI results from the distortion of space under the gravitational field in the radial direction.

Using numerical iteration of Eqs. (1)-(3) we calculated the trajectories of circularly polarized photons (red lines in Fig. 1) in the gravitational field of a subwavelength black hole (blue spheres in Fig. 1) with the Schwarzschild radius $r_0 = 2GM/c^2$. The calculated results clearly show that the incident light is bent by the gravitational field induced by the back hole, and the trajectories are directed to a focal point by the gravitational lens effect.

In addition to the focusing effect, an intriguing phenomenon emerges as a result of the anomalous velocity in Eq. (1). For wavelengths that are much smaller than the size of the black hole, the trajectories are well focused to a single point [Fig. 1(a) and Fig. 2(a)], but for wavelengths larger than the black hole the streamlines remain separated near the focus. This arrangement of the streamlines can be more clearly seen in Fig. 2(b), which shows an enlargement of the region near the focus in Fig. 1(b). The trajectories are chirally aligned – they still preserve the rotational symmetry, while the mirror symmetry is lost. For an incident beam of opposite circular polarizations, the distribution of the trajectories is similar but show different chiralities [Fig. 2(c)].

The chiral distributions of the trajectories clearly demonstrate that the gravitational field induces SOI onto the photons. Because the initial momentum of the incident plane wave is parallel with the z-direction, the photons initially carry no orbital AM, only spin AM, because of their circular polarization. When the photons propagate past the black hole, the gravitational lens effect leads to a redirection of the momentum along the $\theta$ direction (defined in Fig. 1) and leads to the focusing of the light. This focusing behavior does not contribute to an orbital AM. The deviation of the spin AM implies that the total AM momentum changes in the z-direction. However, since this process is elastic, the total AM is conserved and thus, spin AM is transferred to orbital AM via the SOI effect. According to Eq. (2) the anomalous velocity resulting from the SOI is along the azimuthal direction $\phi$ and thus leads to a transverse shift of the trajectories, as revealed by the chiral distribution in Fig. 1(b). This transverse shift in turn leads to an azimuthal component of momentum in Eq. (1) and an orbital AM in the *z*-direction. This orbital AM is the result of SOI of photons in the gravitational field which convert the spin AM to the orbital part. The effect of the SOI is determined by the anomalous velocity, which is proportional to the ratio between the wavelength and the smallest distance of the photon from the center of the black hole as shown in Eq. (2). Therefore, the chiral trajectories resulting from gravitational-field-induced SOI only occur for wavelengths of the incident light that are larger than the Schwarzschild radius [Figs. 1(a) and (b)].

## 3  Discussion and conclusion

The above theoretical analysis shows that: i) the gravitational field around black holes of subwavelength size twist the trajectories of incident circularly polarized photons because of SOI induced by the gravitational field, ii) the resulting chiral trajectories can lead to an observable image rotation of the gravitational lens. The virtual image is obtained from the extended lines of the trajectories, and therefore the bending of the light will permit observers to see an object hidden behind the stars. If the light from the object is circularly polarized, as shown in our calculations (Fig. 1), the light lines will manifest spin-dependent transverse propagation-angle deviations and the object image obtained from the extended lines to be rotated at a transverse angle rotation with respect to object. Considering the axial symmetry of the system, this transverse angle deviation implies a rotation of the image with respect to the object.

The findings open for probing subwavelength black holes via the SOI induced by gravitational lensing on circularly polarized light. Such black holes with mass equal to the earth have a Schwarz radius of only a few millimeter and much smaller black holes have been predicted to emerge during the operation of high energy colliders [22,23].

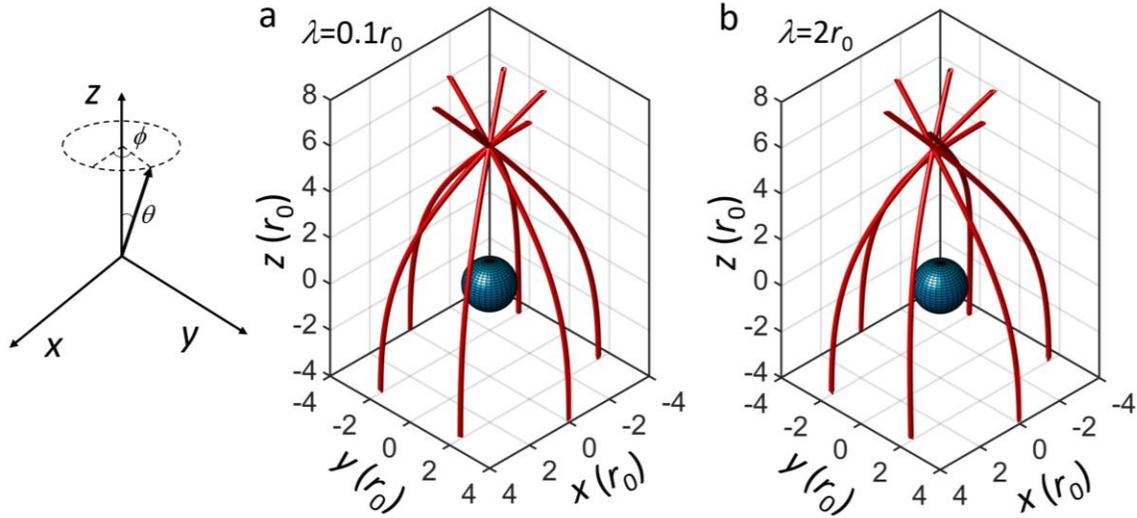

**Fig. 1** Spin-orbit interaction induced by the gravitational field around subwavelength black holes twists the trajectories of incident circularly polarized light. Images show the focusing of light with helicity $\lambda = -\hbar$ in the gravitational lens induced by a blackhole for wavelength of $0.1 r_0$ (**a**) and $2 r_0$ (**b**), where $r_0$ is the Schwarzschild radius of the blackhole (illustrated by the blue sphere). The trajectories were calculated by iterating Eqs. (1)-(3), starting from the initial position at $z=-2 r_0$ and $\sqrt{x^2 + y^2} = 4 r_0$ with the momentum oriented toward z direction.

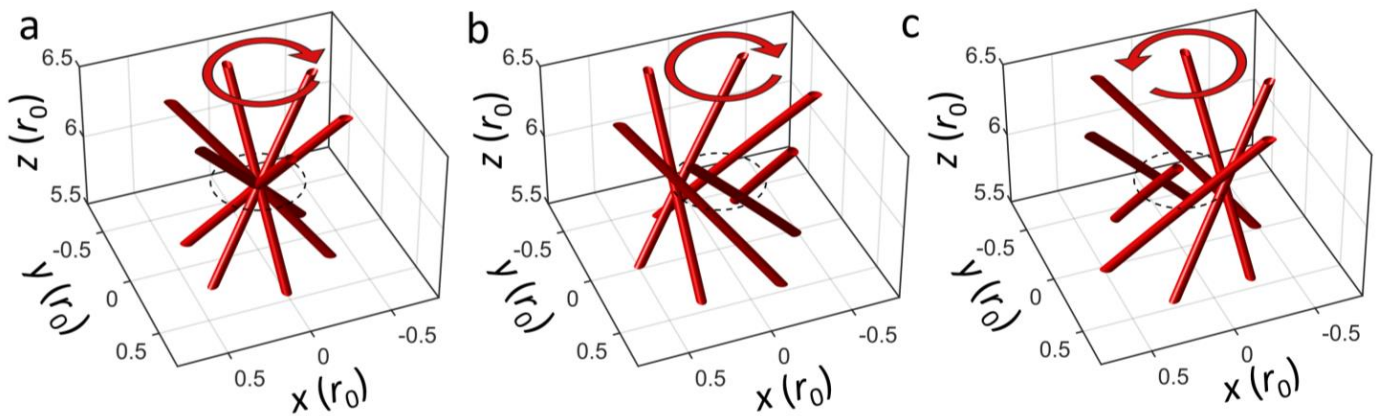

**Fig. 2** The streamline of the momentum near the focus of the gravitational lens for incident light of helicity $\lambda = -\hbar$ for wavelength $0.1 r_0$ (**a**) and $2 r_0$ (**b**), and helicity $\lambda = +\hbar$ for wavelength $2 r_0$

(b), as denoted by the circular arrows. (a) and (b) are enlargements of the focus regions in from Figs. 1(a) and (b).

**References**

1. D. Pan, H. Wei, L. Gao, and H. Xu, Strong Spin-Orbit Interaction of Light in Plasmonic Nanostructures and Nanocircuits, *Phys. Rev. Lett.* 117(16), 166803 (2016)

2. P. J. Wang and J. Zhang, Spin–orbit coupling in Bose–Einstein condensate and degenerate Fermi gases, *Front. Phys.* 9(5), 598-612 (2014)

3. J. K. Wang, W. Yi, and W. Zhang, Two-body physics in quasi-low-dimensional atomic gases under spin–orbit coupling, Front. Phys. 11(3), 118102 (2016)

4. M. Onoda, S. Murakami, and N. Nagaosa, Hall effect of light, *Phys. Rev. Lett.* 93(8), 083901 (2004)

5. Y. Gorodetski, A. Niv, V. Kleiner, and E. Hasman, Observation of the spin-based plasmonic effect in nanoscale structures, *Phys. Rev. Lett.* 101(4), 043903 (2008)

6. K. Y. Bliokh, Y. Gorodetski, V. Kleiner, and E. Hasman, Coriolis effect in optics: unified geometric phase and spin-Hall effect, *Phys. Rev. Lett.* 101(3), 030404 (2008)

7. S. Y. Lee, I. M. Lee, J. Park, S. Oh, W. Lee, K. Y. Kim, and B. Lee, Role of magnetic induction currents in nanoslit excitation of surface plasmon polaritons, *Phys. Rev. Lett.* 108(21), 213907 (2012)

8. F. J. Rodríguez-Fortuño, G. Marino, P. Ginzburg, D. O'Connor, A. Martínez, G. A. Wurtz, A. V. Zayats, Near-field interference for the unidirectional excitation of electromagnetic guided modes, *Science* 340(6130), 328-330 (2013)

9. J. Lin, J. P. B. Mueller, Q. Wang, G. Yuan, N. Antoniou, X. C. Yuan, F. Capasso, Polarization-controlled tunable directional coupling of surface plasmon polaritons, *Science* 340(6130), 331-334 (2013)

10. X. Yin, Z. Ye, J. Rho, Y. Wang, and X. Zhang, Photonic spin Hall effect at metasurfaces, *Science* 339(6126), 1405-1407 (2013)

11. N. Shitrit, I. Yulevich, E. Maguid, D. Ozeri, D. Veksler, V. Kleiner, and E. Hasman, Spin-optical metamaterial route to spin-controlled photonics, *Science* 340(6133), 724-726 (2013)

12. J. Petersen, J. Volz, and A. Rauschenbeutel, Chiral nanophotonic waveguide interface based on spin-orbit interaction of light, *Science* 346(6205), 67-71 (2014)

13. D. O'Connor, P. Ginzburg, F. J. Rodríguez-Fortuño, G. A. Wurtz, and A. V. Zayats, Spin–orbit coupling in surface plasmon scattering by nanostructures. *Nat. Commun.* 5, 5327 (2014).

14. J. B. Pendry, D. Schurig, D. R. Smith, Controlling electromagnetic fields, *Science* 312(5781), 1780-1782 (2006)

15. U. Leonhardt, Optical conformal mapping, Science 312(5781), 1777-1780 (2006)

16. D. A. Genov, S. Zhang, X. Zhang, Mimicking celestial mechanics in metamaterials, Nat. Phys. 5(9), 687-692 (2009)

17. W. X. Jiang, and B. G. Cai, An electromagnetic black hole made of metamaterials, *arXiv* 0910, 2159 (2009)

18. C. Sheng, H. Liu, Y. Wang, S. N. Zhu, and D. A. Genov, Trapping light by mimicking gravitational lensing, *Nat. Photon.* 7(11), 902-906 (2013)

19. C. Sheng, R. Bekenstein, H. Liu, S. Zhu and M. Segev, Wavefront Shaping through Emulated Curved Space in Waveguide Settings, *Nat. Commun.* 7, 10747 (2016)

20. P. Gosselin, A. Bérard A, and H. Mohrbach, Spin Hall effect of photons in a static gravitational field, *Phys. Rev. D* 75(8), 084035 (2007)

21. Y. N. Obukhov, Spin, gravity, and inertia, *Phys. Rev. Lett.* 86(2), 192 (2001)

22. S. Dimopoulos and G. Landsberg, Black holes at the large hadron collider, *Phys. Rev. Lett.* 87(16) 161602 (2001)

23. S. B. Giddings and S. Thomas, High energy colliders as black hole factories: The end of short distance physics, *Phys. Rev. D* 65(5), 056010 (2002)